%% file: SiNMEMSapl.TEX
\documentclass[aip,apl,twocolumn,preprnt,preprintnumbers,amsmath,amssymb,superscriptaddress]{revtex4}

\usepackage{graphicx}
\usepackage{dcolumn}
\usepackage{bm}

\usepackage[a4paper,top=2.5cm,bottom=2cm,left=2.5cm,right=2cm]{geometry}	% page margins

\usepackage{amsmath,amssymb,amsfonts} % standard AMS packages

\usepackage{bm}	% bold symbols in math mode \bm{...}
\renewcommand{\mathbf}{\bm}
\usepackage{dsfont}	% proper mathbb format
\renewcommand{\mathbb}{\mathds}	% redefine \mathbb

\usepackage{mathrsfs} % use \mathscr{} for script letters in math

\usepackage{mathtools} % for proper typesetting of := and =:

\usepackage{graphicx,float}
\usepackage[colorlinks,
linkcolor=red,
citecolor=blue,
urlcolor=red]{hyperref}

\usepackage{wrapfig}	% bunch of stuff to wrap text around images
\newcommand{\fref}[1]{Fig.~\ref{#1}}
\renewcommand{\eqref}[1]{Eq.~\ref{#1}}

\renewcommand{\t}[1]{\mathrm{#1}}

\begin{document}

\preprint{}

\title{A strongly-coupled $\Lambda$-type micromechanical system}

\author{Hajime Okamoto}
%\email[]{okamoto.hajime@lab.ntt.co.jp}
\affiliation{Institute of Condensed Matter Physics, $\acute{E}$cole Polytechnique F$\acute{e}$d$\acute{e}$rale de Lausanne (EPFL), CH-1015 Lausanne, Switzerland}
\affiliation{NTT Basic Research Laboratories, Nippon Telegraph and Telephone Corporation, Atsugi 243-0198, Japan}

\author{Ryan Schilling}
\affiliation{Institute of Condensed Matter Physics, $\acute{E}$cole Polytechnique F$\acute{e}$d$\acute{e}$rale de Lausanne (EPFL), CH-1015 Lausanne, Switzerland}
\author{Hendrik Sch$\ddot{\rm u}$tz}
\affiliation{Institute of Condensed Matter Physics, $\acute{E}$cole Polytechnique F$\acute{e}$d$\acute{e}$rale de Lausanne (EPFL), CH-1015 Lausanne, Switzerland}

\author{Vivishek Sudhir}
\affiliation{Institute of Condensed Matter Physics, $\acute{E}$cole Polytechnique F$\acute{e}$d$\acute{e}$rale de Lausanne (EPFL), CH-1015 Lausanne, Switzerland}

\author{Dalziel J. Wilson}
\affiliation{Institute of Condensed Matter Physics, $\acute{E}$cole Polytechnique F$\acute{e}$d$\acute{e}$rale de Lausanne (EPFL), CH-1015 Lausanne, Switzerland}

\author{Hiroshi Yamaguchi}
\affiliation{NTT Basic Research Laboratories, Nippon Telegraph and Telephone Corporation, Atsugi 243-0198, Japan}

\author{Tobias J. Kippenberg}
%\email[]{tobias.kippenberg@epfl.ch}
\affiliation{Institute of Condensed Matter Physics, $\acute{E}$cole Polytechnique F$\acute{e}$d$\acute{e}$rale de Lausanne (EPFL), CH-1015 Lausanne, Switzerland}

\date{\today}% It is always \today, today,

\begin{abstract}
We study a classical $\Lambda$-type three-level system based on three high-$Q$ micromechanical beam resonators embedded in a gradient electric field.  By modulating the strength of the field at the difference frequency between adjacent beam modes, we realize strong dynamic two-mode coupling, via the dielectric force.  Driving adjacent pairs simultaneously, we observe the formation of a purely mechanical `dark' state and an all-phononic analog of coherent population trapping --- signatures of strong three-mode coupling. The $\Lambda$-type micromechanical system is a natural extention of previously demonstrated `two-level' micromechanical systems and offers new perspectives on the architecture of all-phononic micromechanical circuits and arrays.
\end{abstract}

\maketitle

%\section{Introduction}
The ability to control phonon transport in micromechanical systems has important consequences for applications such as low-power mechanical filters, switches, routers, memories, and logic gates \cite{trigo2002confinement,maldovan2013sound,habraken2012continuous,schmidt2012optomechanical,safavi2011proposal,zhu2007stored,mahboob2008bit,hatanaka2014phonon,sklan2015splash}. As a basic starting point, considerable effort has been aimed towards achieving tunable coupling between two modes of a micromechanical resonator, using a combination of piezoelectric \cite{okamoto2013coherent,mahboob2012phonon}, photothermal \cite{ohta2015optically,liu2015optical}, and dielectric \cite{unterreithmeier2009universal} forces.  Realization of strong coupling ---  that is, coherent energy exchange between two modes at a rate ($g$) larger than their mechanical dissipation ($\gamma$) --- has recently enabled classical analogs of `two-level system' coherent control, including Rabi flops, Ramsey fringes and Hahn echo \cite{okamoto2013coherent,faust2013coherent}. These demonstrations provide a dress rehearsal for phonon transport in future micromechanical circuits and arrays \cite{truitt2007efficient,bargatin2012large}.  

In this Letter, we demonstrate strong coupling between three radiofrequency micromechanical resonators forming a classical analog of a $\Lambda$-type three-level system (two nearly degenerate, low energy levels and single high energy level).   The inclusion of a third `level' opens the door to a rich variety of physics not accessible to two-level systems, such as analogs of EIT (electromagnetically induced transparency \cite{harris2008electromagnetically}) and CPT (coherent population trapping \cite{arimondo1996coherent}). To our knowledge, a fully micromechanical $\Lambda$-type system has not been previously implemented.  By contrast, V-type three-level systems formed by a pair of optical cavity modes coupled to a common mechanical resonator mode have been extensively explored in the context of cavity optomechanics \cite{weis2010optomechanically,dong2012optomechanical,wang2012usinginterference}, enabling demonstrations of optomechanical state transfer \cite{weis2010optomechanically} and optomechanical dark states \cite{dong2012optomechanical}.  In analogy to the latter, we show that by driving both pairs of the $\Lambda$-system micromechanical system simultaneously, destructive interference leads to the formation of a mechanical dark state and a phononic analog of CPT.

\begin{figure}[tr]
	\begin{center}
		\includegraphics[scale=0.87]{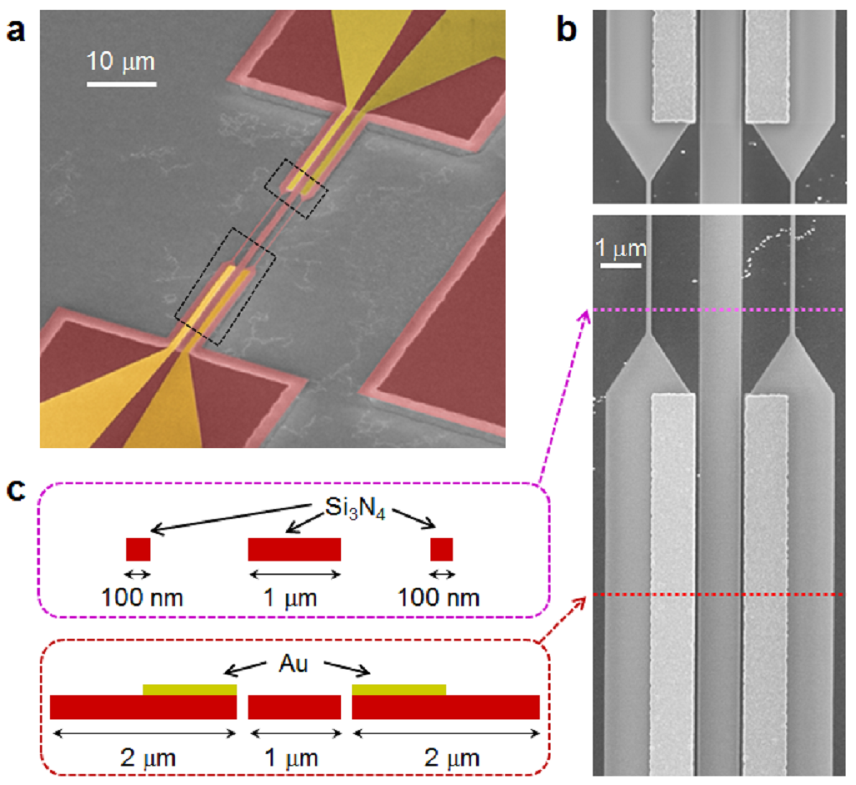}
		\caption{\textbf{Overview of the device.} (a) False-color scanning electron micrograph of the device; yellow, Au; red, Si$_3$N$_4$; grey, Si substrate. (b) Magnified scanning electron micrograph, highlighting the region surrounded by black squares in (a). (c) Two sketches of the sample in cross-section, corresponding to horizontal cuts along the pink and red dotted line in (b). The thickness of Si$_3$N$_4$ and metal layers are 100 and 45 nm, respectively. The horizontal gap between adjacent beams is 80 nm.}
		\label{Fig1}
	\end{center}
	\vspace{-15pt}
\end{figure}

The micromechanical system, shown in \fref{Fig1}, consists of a planar stack of three high-stress Si$_3$N$_4$ microbeams.  The outer left and right beams possess fundamental out-of-plane flexural modes with frequencies $\omega_\t{L} \sim2\pi\cdot1.67$ MHz and $\omega_\t{R} \sim2\pi\cdot1.72$ MHz, respectively.  The middle beam has a fundamental mode frequency of $\omega_\t{M} \sim2\pi\cdot4.50$ MHz.  To couple these modes, a voltage is applied between two electrodes patterned on the outer beams.  The resulting electric field produces a dielectric potential, which, owing to the spatial dependence of the field, gives rise to a static intermodal coupling \cite{faust2013coherent}, analogous to the strain-mediated coupling of two beams sharing a non-rigid anchor \cite{okamoto2013coherent}.  We wish to emphasize that, contrary to the latter case, dielectric intermode coupling can be switched completely off by grounding the electrodes.  In this sense, the mechanical modes may be considered to reside on three physically isolated mechanical resonators.  Though conceptually the same as three modes of a single resonator, physical separation  is advantageous for building micromechanical circuits/arrays, as it enables the resonators to be fictionalized for independent purposes.  For instance, we envision that the central beam in \fref{Fig1} may be integrated into the a near-field of an optical cavity \cite{wilson2015measurement}, for precision readout and actuation.

\begin{figure}[tl!]
	\begin{center}
		\includegraphics[scale=0.87]{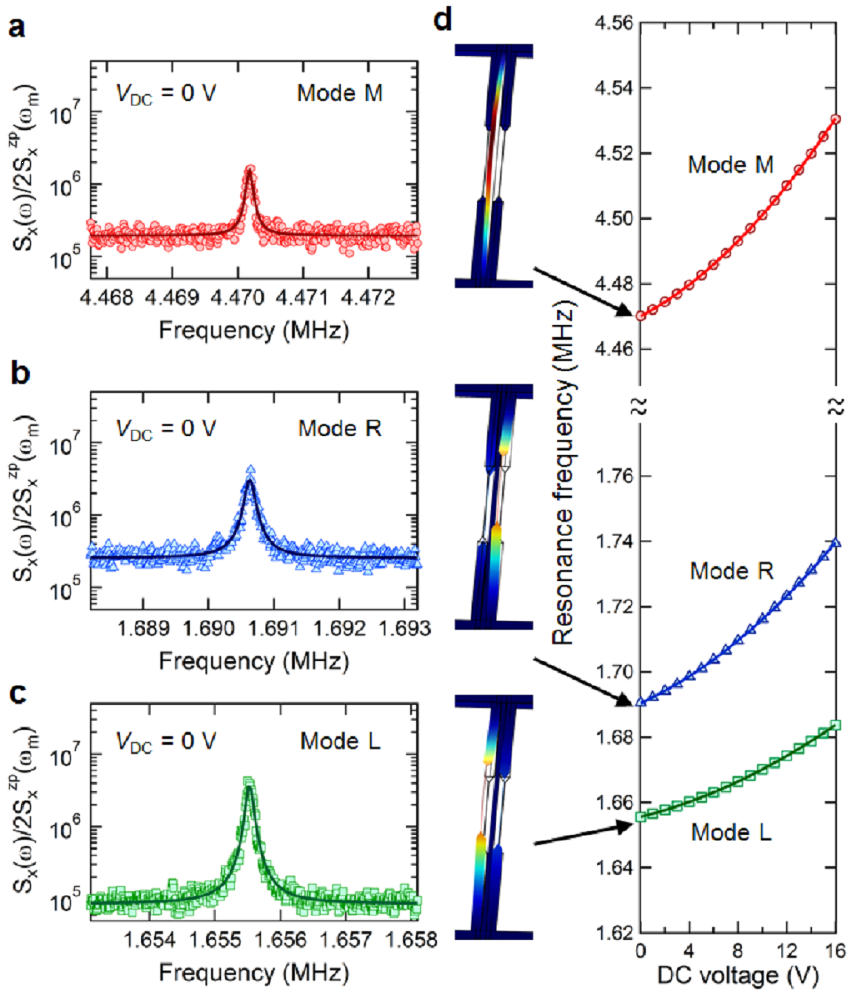}
		\caption{\textbf{Dielectric tuning of mechanical eigenfrequencies and optical readout}. (a-c) Thermal motion of the middle beam (mode M), right beam (mode R), and left beam (mode L), readout using a lensed-fiber interferometer. For these measurements no DC voltage is applied across the metal electrodes. Displacement noise spectral densities, $S_x$, are expressed in units scaled to the twice the peak zero point spectral density, $S_x^{\rm zp}(\omega_\t{m})$ (i.e., the displacement noise produced by single thermal phonon). In these units, the peak spectral density is equal to with the thermal occupation, $n_\t{th}\approx k_\t{B}T/\hbar\omega_\t{m}$, where $\omega_\t{m}$ is the mechanical frequency.  (d) Dependence of mode frequencies on the DC voltage applied across the electronics. Mode shapes computed by the finite element method are shown on the left.}
		\label{Fig2}
	\end{center}
	\vspace{-20pt}
\end{figure}

Applying a DC voltage produces static coupling between the beams; however, energy transfer between vibrational modes remains weak because of their different eigenfrequencies.  To overcome this non-degeneracy, an AC voltage may be applied at the difference frequency between modes of adjacent beams.  The resulting modulated static coupling strength (see \eqref{Eq1}) leads to energy flow between the two modes --- hereafter referred to as dynamic intermode coupling.  The magnitude of the dynamic coupling rate is proportional to the product of the AC voltage, the DC voltage, and a geometric factor proportional to the cross derivative of the dielectric potential.  When sufficiently large, the dynamically coupled modes exhibit normal mode splitting.  The magnitude of this splitting corresponds to their coupling rate.  In our system, owing to the high room temperature quality factor of the tensily stressed beams $(Q\sim 10^4)$, the normal mode splitting can be made much larger than that of mechanical dissipation (i.e. energy decay) rate, corresponding to ``strong coupling".

Experimental characterization of the device is shown in \fref{Fig2}. Mechanical spectra are recorded used a lensed fiber-based interferometer \cite{azak2007nanomechanical}  with a spot size ($\sim4\,\mu$m) large enough to simultaneously record the thermal displacement of each beam. The optical field, supplied by a 780 nm Ti:sapphire laser, is attenuated until photothermal effects are negligible. Sample and fiber are embedded in a vacuum chamber at $10^{-4}$ mbar.  When no voltage is applied across the electrodes, the fundamental out-of-plane modes of the middle (M), right (R), and left (L) beams reside at their natural frequencies with energy dissipation rates of $\gamma_\t{M} = 2\pi\cdot 100$ Hz, $\gamma_\t{R} = 2\pi\cdot170$ Hz, and $\gamma_\t{L} = 2\pi\cdot 140$ Hz, respectively, shown in descending order in \fref{Fig2}a-c.  The identity of each mode is verified by finite element simulation (using COMSOL, see \fref{Fig2}d).  When a DC voltage is applied to the electrodes, the mechanical frequencies are blue-shifted (\fref{Fig2}d).  The frequency shift is nearly quadratic, consistent with the dielectric force in a linear field gradient \cite{unterreithmeier2009universal}.  

\begin{figure}[tr]
	\begin{center}
		\includegraphics[scale=0.87]{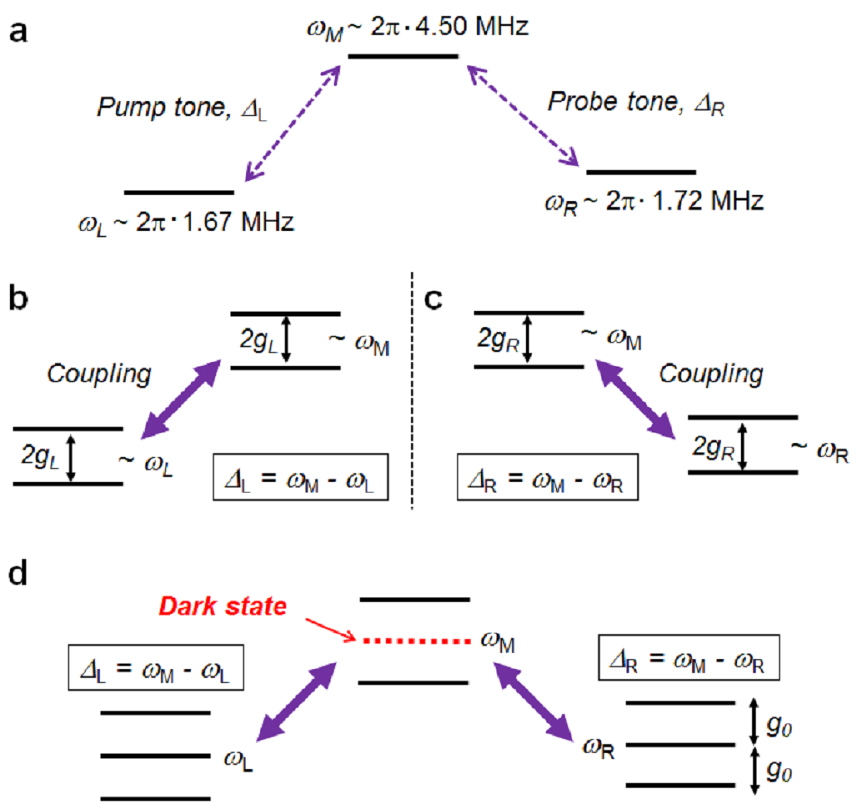}
		\caption{\textbf{Energy level diagram.} (a) Energy level diagram of the $\Lambda$-type micromechanical system in the absence of dynamic coupling. (b,c) Energy level diagram in the presence of pump (probe) tone $\Delta_\t{L(R)} = \omega_\t{M} - \omega_\t{L(R)}$, exhibiting normal mode splitting of the left (right) and middle beam modes. (d) Energy level diagram of the three-level system in the presence of simultaneous pump and probe tones.  All three levels exhibit mode splitting.  Notably, one of the dressed states is `dark' in the sense that involves no motion of the center beam.}
		\label{Fig3}
	\end{center}
	\vspace{-15pt}
\end{figure}

\begin{figure}[tl!]
	\vspace{-5pt}
	\begin{center}
		\includegraphics[scale=0.9]{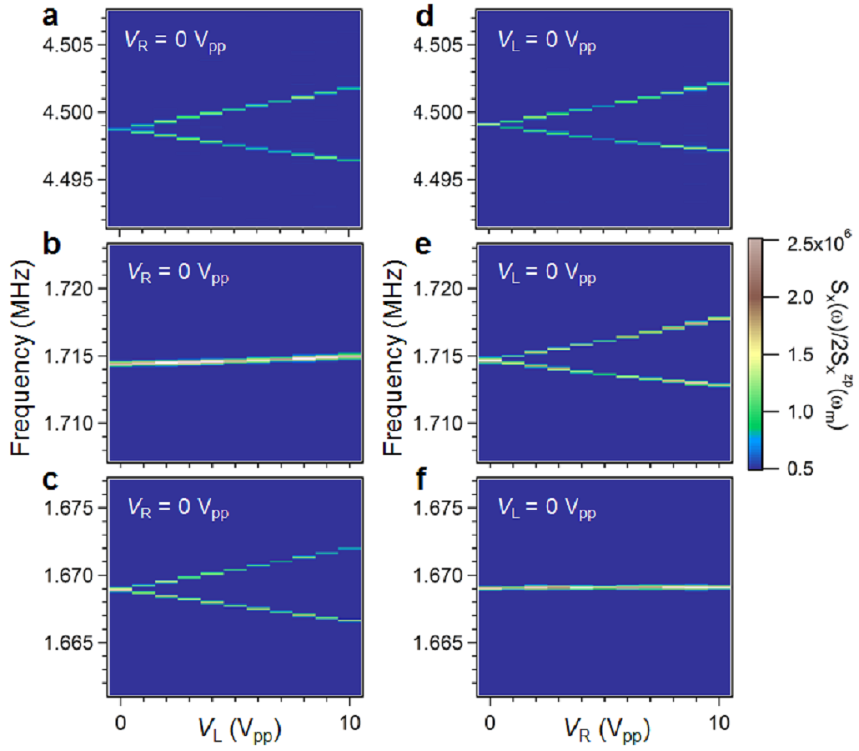}
		\caption{\textbf{Dynamic intermode coupling by dielectric frequency modulation}. Left column: Noise spectrum of mode M (a), mode R (b), and mode L (c) versus pump amplitude ($V_\t{L}$) with pump frequency tuned to $\Delta_\t{L} = \omega_\t{M} - \omega_\t{L}$, $V_\t{DC} =$ 10 V, and no probe voltage applied. Right column: Noise spectrum of mode M (d), mode R (e), and mode L (f) versus probe amplitude (V$_\t{p}$) with probe frequency tuned to $\Delta_\t{R} = \omega_\t{M} - \omega_\t{R}$, $V_\t{DC} =$ 10 V, and no pump voltage applied.}
		\label{Fig4}
	\end{center}
	\vspace{-20pt}
\end{figure}

To realize dynamic intermode coupling, an AC voltage is applied across the electrodes at one of the two difference frequencies, $\Delta _\t{L(R)}=\omega_\t{M}-\omega_\t{L(R)}$.  (Following \cite{tian2012adiabatic,wang2012usingdarkmodes,wang2012usinginterference}, we refer to these as the `pump' ($\Delta_\t{L}$) and `probe' ($\Delta_\t{R}$) tones. See \fref{Fig3}.)  For sufficiently large AC voltages, normal mode splitting is observed in the thermal displacement noise spectrum.  Shown in \fref{Fig4} is the displacement noise spectrum plotted versus AC voltage amplitude $V_\t{L(R)}$ for a fixed DC offset of $V_\t{DC} = 10$ V.  The magnitude of the mode splitting is proportional to the AC voltage amplitude, and corresponds to twice the energy coupling rate $g_\t{L(R)}$ between the left(right) and middle beam modes. Strong coupling, $g_\t{L(R)}>\{\gamma_\t{L(R)},\gamma_\t{M}\}$, is achieved for $V_\t{L(R)} >$ 1 $V_\t{pp}$ in both cases.  At the largest AC voltage amplitudes, the cooperativities achieved are $C_\t{L(R)} = 4g_\t{L(R)}^2/(\gamma_\t{L(R)}\gamma_\t{M}) > 1 \times 10^4$.

\begin{figure*}[t!]
\begin{center}
\includegraphics[scale=0.9]{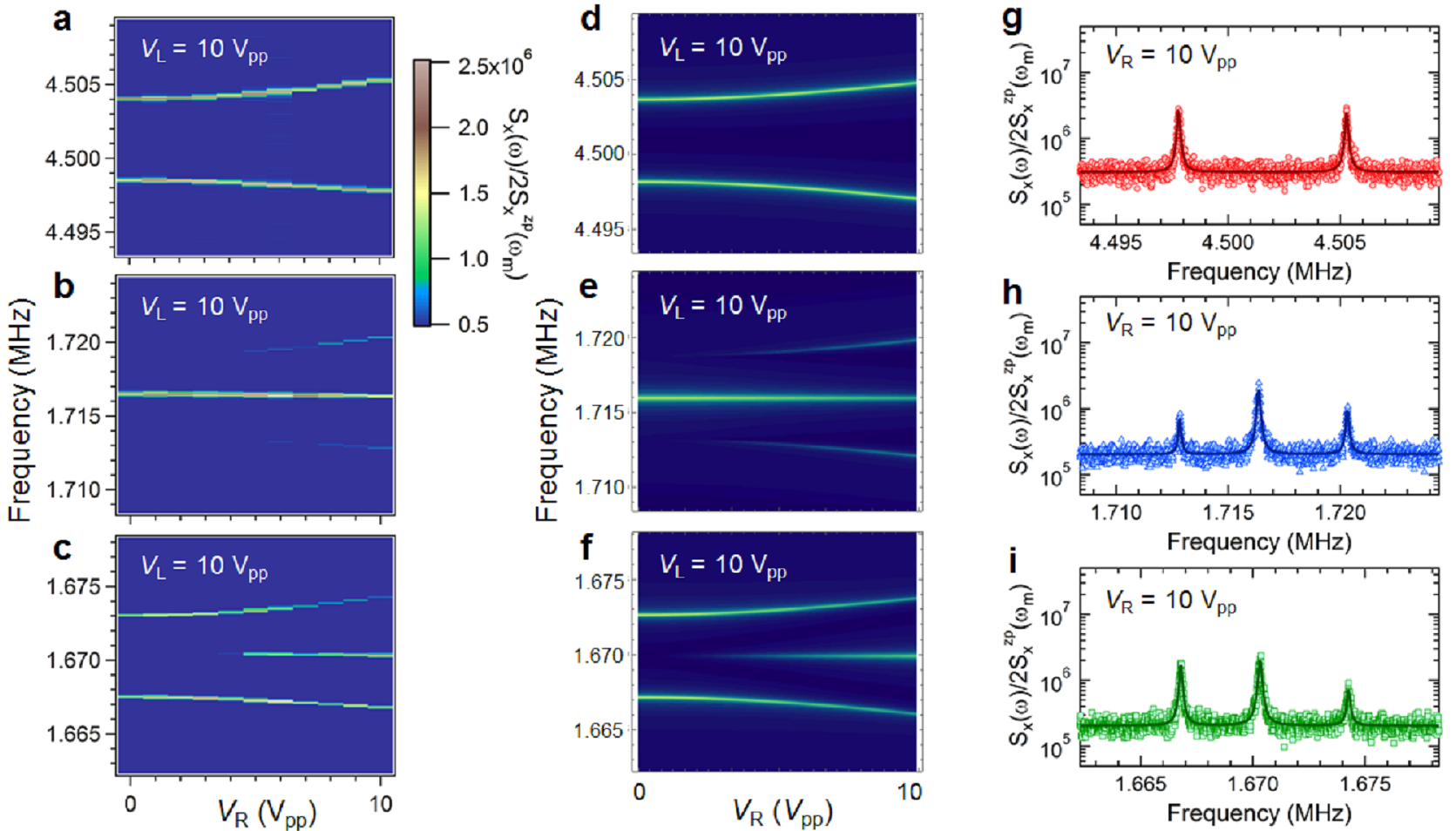}
\caption{\textbf{Analog of coherent population trapping among three dynamically coupled micromechanical beams.} Probe voltage ($V_\t{R}$) dependence of mode M (a), mode R (b), and mode L (c) for $V_\t{DC} =$ 10 V when the pump frequency is tuned to $\Delta_\t{L} = \omega_\t{M} - \omega_\t{L}$ and the probe frequency is tuned to $\Delta_\t{R} = \omega_\t{M} - \omega_\t{R}$. The pump voltage is here fixed at $V_\t{L} =$ 10 $V_\t{pp}$. Simulated results for (a), (b), and (c) are shown in (d), (e), (f), respectively. Cross-sections of (a), (b), and (c) at $V_\t{R} =10\;V_\t{pp}$ are shown in (g), (h), and (i).}
\label{Fig5}
\end{center}
\vspace{-25pt}
\end{figure*}

When the pump and probe tones are applied simultaneously, strong coupling is induced among all three beams. This coupling is akin to a resonant Raman interaction, and features the formation of a dynamically decoupled `dark' state due to destructive interference (see \fref{Fig3}d).  To study this effect, the displacement noise spectrum is recorded versus probe strength ($V_\t{R}$) for a fixed pump strength of $V_\t{L} = 10\, \t{V}_\t{pp}$ and a fixed DC offset of $V_\t{DC}=10$ V.  As the probe strength is increased, a trio of dressed states emerges near $\omega_\t{R}$ (\fref{Fig5}b) and $\omega_\t{L}$ (\fref{Fig5}c), reflecting the onset of strong dynamic coupling between all three beams. Notably, only two peaks appear near $\omega_\t{M}$ even when the three mechanical resonators are strongly coupled (\fref{Fig5}a).  The energetically allowed third peak appears `dark' because of destructive interference  between modes L, R, and M.  This effect is analogous to coherent population trapping, as explored in optomechanical systems \cite{wang2012usinginterference,tian2012adiabatic,wang2012usingdarkmodes}.

The experimental results shown in \fref{Fig5}a-c are reproduced in \fref{Fig5}d-f using a classical coupled-resonator model: 
\begin{subequations}\label{Eq1}
\begin{align}
\ddot{x}_\t{L} + \gamma_\t{L}\dot{x}_\t{L} + \omega_\t{L}^2 x_\t{L} &= F_\t{L} - g_\t{L}x_\t{M} \\
\ddot{x}_\t{M} + \gamma_\t{M}\dot{x}_\t{M} + \omega_\t{M}^2 x_\t{M} &= F_\t{M} - g_\t{L}x_\t{L} - g_\t{R}x_\t{R} \\
\ddot{x}_\t{R} + \gamma_\t{R}\dot{x}_\t{R} + \omega_\t{R}^2 x_\t{R} &= F_\t{R} - g_\t{R}x_\t{M}.
\end{align}
\end{subequations}
Here $x_\t{L(R,M)}$ is the generalized displacement of the left(right,middle) beam mode, $g_\t{L(R)} = \Omega_\t{L(R)}\cos(\Delta_\t{L(R)}t)$ is the modulated intermode coupling rate with strength $\Omega_\t{L(R)}$ and $F_\t{L(R,M)}$ is a generalized external force. The normal modes of \eqref{Eq1} are: $x_\pm = (g_\t{L}x_\t{L} + g_\t{R}x_\t{R} \pm g_\t{0}x_\t{M})/(\sqrt{2}g_0)$ and $x_\t{D} = (-g_\t{R}x_\t{L} + g_\t{L}x_\t{R})/g_0$, where $g_0 = \sqrt{g_\t{L}^2 + g_\t{R}^2}$. $x_\t{D}$ corresponds to the `dark' mode, which contains no contribution from $x_\t{M}$. To obtain \fref{Fig5}d-f we solved \eqref{Eq1} numerically using the rotating wave approximation \cite{okamoto2013coherent}, viz., assuming
\begin{equation}
x_\t{L}(t) = \t{Re} \left[ \sum^\infty_{m,n=-\infty}a_{m,n}(t)e^{i(\omega+m\Delta_\t{L} + n\Delta_\t{R})t} \right]
\label{Eq2}
\end{equation}
(and similar expressions for $x_\t{R}$ and $x_\t{M}$).  Models in \fref{Fig5}d-f are produced using the first-order approximation $m=n=1$.
\\

In summary, we have demonstrated strong dynamic coupling of three Si$_3$N$_4$ micromechanical resonators via dielectric forces, enabling formation of a mechanical `dark' state and a classical analog of coherent phonon population trapping.  It bears emphasis that experiments were performed at room temperature and that the device (included mechanical and electrical elements) is fully integrated on a chip, allowing for robust control. Pulsed switching of the gradient field is available, thus two-level control techniques (albeit here in the entirely classical analog) such as Rabi oscillation and Ramsey fringe measurements \cite{okamoto2013coherent,faust2013coherent}) and three-level techniques like pulsed EIT \cite{tian2012adiabatic} are available. This control might be extended, for example, to nodes of micromechanical resonator circuits and arrays \cite{truitt2007efficient,bargatin2012large}. % and phonon waveguides \cite{hatanaka2014phonon}. %, thus it is compatible with integrated micromechanical systems. 
We also remark that, by design, the three-beam system shown in \fref{Fig1} is readily incorporated in the evanescent near-field of a optical microdisk cavity; this platform has been shown to enable state-of-the-art interferometric displacement sensitivities as well as efficient radiation pressure force actuation \cite{gavartin2012hybrid,wilson2015measurement}, offering compelling opportunities for coherent coupling of electrical fields and optical fields \cite{bagci2014optical,regal2011cavity,andrews2014bidirectional,bochmann2013nanomechanical}.
\linebreak

%%\section{Summary}

%This enables on-demand control of phonon transfer between independent mechanical systems, which permits the development of all-phononic micromechanical circuits.

%\section{Acknowledgements}

We thank NTT for granting and supporting the sabbatical leave of H.O.. We acknowledge nanofabrication advice from M. Zervas and A. Feofanov. The sample was fabricated at the CMi (Center for Micro-Nanotechnology) at EPFL. Research was funded by an ERC Advanced Grant (QuREM), the Marie Curie Initial Training Network for Cavity Quantum Optomechanics (cQOM), and partial support from MEXT KAKENHI (Grant No. 15H05869 and 15K21727). D.J.W. acknowledges support from the European Commission through a Marie Skodowska-Curie Fellowship (IIF project 331985).

\bibliographystyle{apsrev4-1}
\input{ref1.bbl}
%\bibliography{ref1}

\end{document}

%% file: ref1.bbl

%% file: SiNMEMSapl.bbl
\begin{thebibliography}{31}%
\makeatletter
\providecommand \@ifxundefined [1]{%
 \@ifx{#1\undefined}
}%
\providecommand \@ifnum [1]{%
 \ifnum #1\expandafter \@firstoftwo
 \else \expandafter \@secondoftwo
 \fi
}%
\providecommand \@ifx [1]{%
 \ifx #1\expandafter \@firstoftwo
 \else \expandafter \@secondoftwo
 \fi
}%
\providecommand \natexlab [1]{#1}%
\providecommand \enquote  [1]{``#1''}%
\providecommand \bibnamefont  [1]{#1}%
\providecommand \bibfnamefont [1]{#1}%
\providecommand \citenamefont [1]{#1}%
\providecommand \href@noop [0]{\@secondoftwo}%
\providecommand \href [0]{\begingroup \@sanitize@url \@href}%
\providecommand \@href[1]{\@@startlink{#1}\@@href}%
\providecommand \@@href[1]{\endgroup#1\@@endlink}%
\providecommand \@sanitize@url [0]{\catcode `\\12\catcode `\$12\catcode
  `\&12\catcode `\#12\catcode `\^12\catcode `\_12\catcode `\%12\relax}%
\providecommand \@@startlink[1]{}%
\providecommand \@@endlink[0]{}%
\providecommand \url  [0]{\begingroup\@sanitize@url \@url }%
\providecommand \@url [1]{\endgroup\@href {#1}{\urlprefix }}%
\providecommand \urlprefix  [0]{URL }%
\providecommand \Eprint [0]{\href }%
\providecommand \doibase [0]{http://dx.doi.org/}%
\providecommand \selectlanguage [0]{\@gobble}%
\providecommand \bibinfo  [0]{\@secondoftwo}%
\providecommand \bibfield  [0]{\@secondoftwo}%
\providecommand \translation [1]{[#1]}%
\providecommand \BibitemOpen [0]{}%
\providecommand \bibitemStop [0]{}%
\providecommand \bibitemNoStop [0]{.\EOS\space}%
\providecommand \EOS [0]{\spacefactor3000\relax}%
\providecommand \BibitemShut  [1]{\csname bibitem#1\endcsname}%
\let\auto@bib@innerbib\@empty
%</preamble>
\bibitem [{\citenamefont {Trigo}\ \emph {et~al.}(2002)\citenamefont {Trigo},
  \citenamefont {Bruchhausen}, \citenamefont {Fainstein}, \citenamefont
  {Jusserand},\ and\ \citenamefont {Thierry-Mieg}}]{trigo2002confinement}%
  \BibitemOpen
  \bibfield  {author} {\bibinfo {author} {\bibfnamefont {M.}~\bibnamefont
  {Trigo}}, \bibinfo {author} {\bibfnamefont {A.}~\bibnamefont {Bruchhausen}},
  \bibinfo {author} {\bibfnamefont {A.}~\bibnamefont {Fainstein}}, \bibinfo
  {author} {\bibfnamefont {B.}~\bibnamefont {Jusserand}}, \ and\ \bibinfo
  {author} {\bibfnamefont {V.}~\bibnamefont {Thierry-Mieg}},\ }\href
  {http://journals.aps.org/prl/pdf/10.1103/PhysRevLett.89.227402} {\bibfield
  {journal} {\bibinfo  {journal} {Phys. Rev. Lett.}\ }\textbf {\bibinfo
  {volume} {89}},\ \bibinfo {pages} {227402} (\bibinfo {year}
  {2002})}\BibitemShut {NoStop}%
\bibitem [{\citenamefont {Maldovan}(2013)}]{maldovan2013sound}%
  \BibitemOpen
  \bibfield  {author} {\bibinfo {author} {\bibfnamefont {M.}~\bibnamefont
  {Maldovan}},\ }\href
  {http://www.nature.com/nature/journal/v503/n7475/abs/nature12608.html}
  {\bibfield  {journal} {\bibinfo  {journal} {Nature}\ }\textbf {\bibinfo
  {volume} {503}},\ \bibinfo {pages} {209} (\bibinfo {year}
  {2013})}\BibitemShut {NoStop}%
\bibitem [{\citenamefont {Habraken}\ \emph {et~al.}(2012)\citenamefont
  {Habraken}, \citenamefont {Stannigel}, \citenamefont {Lukin}, \citenamefont
  {Zoller},\ and\ \citenamefont {Rabl}}]{habraken2012continuous}%
  \BibitemOpen
  \bibfield  {author} {\bibinfo {author} {\bibfnamefont {S.}~\bibnamefont
  {Habraken}}, \bibinfo {author} {\bibfnamefont {K.}~\bibnamefont {Stannigel}},
  \bibinfo {author} {\bibfnamefont {M.~D.}\ \bibnamefont {Lukin}}, \bibinfo
  {author} {\bibfnamefont {P.}~\bibnamefont {Zoller}}, \ and\ \bibinfo {author}
  {\bibfnamefont {P.}~\bibnamefont {Rabl}},\ }\href
  {http://iopscience.iop.org/article/10.1088/1367-2630/14/11/115004/pdf}
  {\bibfield  {journal} {\bibinfo  {journal} {New J. Phys.}\ }\textbf {\bibinfo
  {volume} {14}},\ \bibinfo {pages} {115004} (\bibinfo {year}
  {2012})}\BibitemShut {NoStop}%
\bibitem [{\citenamefont {Schmidt}\ \emph {et~al.}(2012)\citenamefont
  {Schmidt}, \citenamefont {Ludwig},\ and\ \citenamefont
  {Marquardt}}]{schmidt2012optomechanical}%
  \BibitemOpen
  \bibfield  {author} {\bibinfo {author} {\bibfnamefont {M.}~\bibnamefont
  {Schmidt}}, \bibinfo {author} {\bibfnamefont {M.}~\bibnamefont {Ludwig}}, \
  and\ \bibinfo {author} {\bibfnamefont {F.}~\bibnamefont {Marquardt}},\ }\href
  {http://iopscience.iop.org/article/10.1088/1367-2630/14/12/125005/meta}
  {\bibfield  {journal} {\bibinfo  {journal} {New J. Phys.}\ }\textbf {\bibinfo
  {volume} {14}},\ \bibinfo {pages} {125005} (\bibinfo {year}
  {2012})}\BibitemShut {NoStop}%
\bibitem [{\citenamefont {Safavi-Naeini}\ and\ \citenamefont
  {Painter}(2011)}]{safavi2011proposal}%
  \BibitemOpen
  \bibfield  {author} {\bibinfo {author} {\bibfnamefont {A.~H.}\ \bibnamefont
  {Safavi-Naeini}}\ and\ \bibinfo {author} {\bibfnamefont {O.}~\bibnamefont
  {Painter}},\ }\href
  {http://iopscience.iop.org/article/10.1088/1367-2630/13/1/013017/meta}
  {\bibfield  {journal} {\bibinfo  {journal} {New J. Phys.}\ }\textbf {\bibinfo
  {volume} {13}},\ \bibinfo {pages} {013017} (\bibinfo {year}
  {2011})}\BibitemShut {NoStop}%
\bibitem [{\citenamefont {Zhu}\ \emph {et~al.}(2007)\citenamefont {Zhu},
  \citenamefont {Gauthier},\ and\ \citenamefont {Boyd}}]{zhu2007stored}%
  \BibitemOpen
  \bibfield  {author} {\bibinfo {author} {\bibfnamefont {Z.}~\bibnamefont
  {Zhu}}, \bibinfo {author} {\bibfnamefont {D.~J.}\ \bibnamefont {Gauthier}}, \
  and\ \bibinfo {author} {\bibfnamefont {R.~W.}\ \bibnamefont {Boyd}},\ }\href
  {http://www.sciencemag.org/content/318/5857/1748.short} {\bibfield  {journal}
  {\bibinfo  {journal} {Science}\ }\textbf {\bibinfo {volume} {318}},\ \bibinfo
  {pages} {1748} (\bibinfo {year} {2007})}\BibitemShut {NoStop}%
\bibitem [{\citenamefont {Mahboob}\ and\ \citenamefont
  {Yamaguchi}(2008)}]{mahboob2008bit}%
  \BibitemOpen
  \bibfield  {author} {\bibinfo {author} {\bibfnamefont {I.}~\bibnamefont
  {Mahboob}}\ and\ \bibinfo {author} {\bibfnamefont {H.}~\bibnamefont
  {Yamaguchi}},\ }\href
  {http://www.nature.com/nnano/journal/v3/n5/abs/nnano.2008.84.html} {\bibfield
   {journal} {\bibinfo  {journal} {Nat. Nano.}\ }\textbf {\bibinfo {volume}
  {3}},\ \bibinfo {pages} {275} (\bibinfo {year} {2008})}\BibitemShut {NoStop}%
\bibitem [{\citenamefont {Hatanaka}\ \emph {et~al.}(2014)\citenamefont
  {Hatanaka}, \citenamefont {Mahboob}, \citenamefont {Onomitsu},\ and\
  \citenamefont {Yamaguchi}}]{hatanaka2014phonon}%
  \BibitemOpen
  \bibfield  {author} {\bibinfo {author} {\bibfnamefont {D.}~\bibnamefont
  {Hatanaka}}, \bibinfo {author} {\bibfnamefont {I.}~\bibnamefont {Mahboob}},
  \bibinfo {author} {\bibfnamefont {K.}~\bibnamefont {Onomitsu}}, \ and\
  \bibinfo {author} {\bibfnamefont {H.}~\bibnamefont {Yamaguchi}},\ }\href
  {http://www.nature.com/nnano/journal/v9/n7/abs/nnano.2014.107.html}
  {\bibfield  {journal} {\bibinfo  {journal} {Nat. Nano.}\ }\textbf {\bibinfo
  {volume} {9}},\ \bibinfo {pages} {520} (\bibinfo {year} {2014})}\BibitemShut
  {NoStop}%
\bibitem [{\citenamefont {Sklan}(2015)}]{sklan2015splash}%
  \BibitemOpen
  \bibfield  {author} {\bibinfo {author} {\bibfnamefont {S.~R.}\ \bibnamefont
  {Sklan}},\ }\href
  {http://scitation.aip.org/content/aip/journal/adva/5/5/10.1063/1.4919584}
  {\bibfield  {journal} {\bibinfo  {journal} {AIP Advances}\ }\textbf {\bibinfo
  {volume} {5}},\ \bibinfo {pages} {053302} (\bibinfo {year}
  {2015})}\BibitemShut {NoStop}%
\bibitem [{\citenamefont {Okamoto}\ \emph {et~al.}(2013)\citenamefont
  {Okamoto}, \citenamefont {Gourgout}, \citenamefont {Chang}, \citenamefont
  {Onomitsu}, \citenamefont {Mahboob}, \citenamefont {Chang},\ and\
  \citenamefont {Yamaguchi}}]{okamoto2013coherent}%
  \BibitemOpen
  \bibfield  {author} {\bibinfo {author} {\bibfnamefont {H.}~\bibnamefont
  {Okamoto}}, \bibinfo {author} {\bibfnamefont {A.}~\bibnamefont {Gourgout}},
  \bibinfo {author} {\bibfnamefont {C.-Y.}\ \bibnamefont {Chang}}, \bibinfo
  {author} {\bibfnamefont {K.}~\bibnamefont {Onomitsu}}, \bibinfo {author}
  {\bibfnamefont {I.}~\bibnamefont {Mahboob}}, \bibinfo {author} {\bibfnamefont
  {E.~Y.}\ \bibnamefont {Chang}}, \ and\ \bibinfo {author} {\bibfnamefont
  {H.}~\bibnamefont {Yamaguchi}},\ }\href
  {http://www.nature.com/nphys/journal/v9/n8/abs/nphys2665.html} {\bibfield
  {journal} {\bibinfo  {journal} {Nat. Phys.}\ }\textbf {\bibinfo {volume}
  {9}},\ \bibinfo {pages} {480} (\bibinfo {year} {2013})}\BibitemShut {NoStop}%
\bibitem [{\citenamefont {Mahboob}\ \emph {et~al.}(2012)\citenamefont
  {Mahboob}, \citenamefont {Nishiguchi}, \citenamefont {Okamoto},\ and\
  \citenamefont {Yamaguchi}}]{mahboob2012phonon}%
  \BibitemOpen
  \bibfield  {author} {\bibinfo {author} {\bibfnamefont {I.}~\bibnamefont
  {Mahboob}}, \bibinfo {author} {\bibfnamefont {K.}~\bibnamefont {Nishiguchi}},
  \bibinfo {author} {\bibfnamefont {H.}~\bibnamefont {Okamoto}}, \ and\
  \bibinfo {author} {\bibfnamefont {H.}~\bibnamefont {Yamaguchi}},\ }\href
  {http://www.nature.com/nphys/journal/v8/n5/abs/nphys2277.html} {\bibfield
  {journal} {\bibinfo  {journal} {Nat. Phys.}\ }\textbf {\bibinfo {volume}
  {8}},\ \bibinfo {pages} {387} (\bibinfo {year} {2012})}\BibitemShut {NoStop}%
\bibitem [{\citenamefont {Ohta}\ \emph {et~al.}(2015)\citenamefont {Ohta},
  \citenamefont {Okamoto}, \citenamefont {Hey}, \citenamefont {Friedland},\
  and\ \citenamefont {Yamaguchi}}]{ohta2015optically}%
  \BibitemOpen
  \bibfield  {author} {\bibinfo {author} {\bibfnamefont {R.}~\bibnamefont
  {Ohta}}, \bibinfo {author} {\bibfnamefont {H.}~\bibnamefont {Okamoto}},
  \bibinfo {author} {\bibfnamefont {R.}~\bibnamefont {Hey}}, \bibinfo {author}
  {\bibfnamefont {K.}~\bibnamefont {Friedland}}, \ and\ \bibinfo {author}
  {\bibfnamefont {H.}~\bibnamefont {Yamaguchi}},\ }\href
  {http://scitation.aip.org/content/aip/journal/apl/107/9/10.1063/1.4930149}
  {\bibfield  {journal} {\bibinfo  {journal} {App. Phys. Lett.}\ }\textbf
  {\bibinfo {volume} {107}},\ \bibinfo {pages} {091906} (\bibinfo {year}
  {2015})}\BibitemShut {NoStop}%
\bibitem [{\citenamefont {Liu}\ \emph {et~al.}(2015)\citenamefont {Liu},
  \citenamefont {Kim},\ and\ \citenamefont {Lauhon}}]{liu2015optical}%
  \BibitemOpen
  \bibfield  {author} {\bibinfo {author} {\bibfnamefont {C.-H.}\ \bibnamefont
  {Liu}}, \bibinfo {author} {\bibfnamefont {I.~S.}\ \bibnamefont {Kim}}, \ and\
  \bibinfo {author} {\bibfnamefont {L.~J.}\ \bibnamefont {Lauhon}},\ }\href
  {http://pubs.acs.org/doi/abs/10.1021/acs.nanolett.5b02586} {\bibfield
  {journal} {\bibinfo  {journal} {Nano Lett.}\ }\textbf {\bibinfo {volume}
  {15}},\ \bibinfo {pages} {6727} (\bibinfo {year} {2015})}\BibitemShut
  {NoStop}%
\bibitem [{\citenamefont {Unterreithmeier}\ \emph {et~al.}(2009)\citenamefont
  {Unterreithmeier}, \citenamefont {Weig},\ and\ \citenamefont
  {Kotthaus}}]{unterreithmeier2009universal}%
  \BibitemOpen
  \bibfield  {author} {\bibinfo {author} {\bibfnamefont {Q.~P.}\ \bibnamefont
  {Unterreithmeier}}, \bibinfo {author} {\bibfnamefont {E.~M.}\ \bibnamefont
  {Weig}}, \ and\ \bibinfo {author} {\bibfnamefont {J.~P.}\ \bibnamefont
  {Kotthaus}},\ }\href
  {http://www.nature.com/nature/journal/v458/n7241/abs/nature07932.html}
  {\bibfield  {journal} {\bibinfo  {journal} {Nature}\ }\textbf {\bibinfo
  {volume} {458}},\ \bibinfo {pages} {1001} (\bibinfo {year}
  {2009})}\BibitemShut {NoStop}%
\bibitem [{\citenamefont {Faust}\ \emph {et~al.}(2013)\citenamefont {Faust},
  \citenamefont {Rieger}, \citenamefont {Seitner}, \citenamefont {Kotthaus},\
  and\ \citenamefont {Weig}}]{faust2013coherent}%
  \BibitemOpen
  \bibfield  {author} {\bibinfo {author} {\bibfnamefont {T.}~\bibnamefont
  {Faust}}, \bibinfo {author} {\bibfnamefont {J.}~\bibnamefont {Rieger}},
  \bibinfo {author} {\bibfnamefont {M.~J.}\ \bibnamefont {Seitner}}, \bibinfo
  {author} {\bibfnamefont {J.~P.}\ \bibnamefont {Kotthaus}}, \ and\ \bibinfo
  {author} {\bibfnamefont {E.~M.}\ \bibnamefont {Weig}},\ }\href
  {http://www.nature.com/nphys/journal/v9/n8/abs/nphys2666.html} {\bibfield
  {journal} {\bibinfo  {journal} {Nat. Phys.}\ }\textbf {\bibinfo {volume}
  {9}},\ \bibinfo {pages} {485} (\bibinfo {year} {2013})}\BibitemShut {NoStop}%
\bibitem [{\citenamefont {Truitt}\ \emph {et~al.}(2007)\citenamefont {Truitt},
  \citenamefont {Hertzberg}, \citenamefont {Huang}, \citenamefont {Ekinci},\
  and\ \citenamefont {Schwab}}]{truitt2007efficient}%
  \BibitemOpen
  \bibfield  {author} {\bibinfo {author} {\bibfnamefont {P.~A.}\ \bibnamefont
  {Truitt}}, \bibinfo {author} {\bibfnamefont {J.~B.}\ \bibnamefont
  {Hertzberg}}, \bibinfo {author} {\bibfnamefont {C.}~\bibnamefont {Huang}},
  \bibinfo {author} {\bibfnamefont {K.~L.}\ \bibnamefont {Ekinci}}, \ and\
  \bibinfo {author} {\bibfnamefont {K.~C.}\ \bibnamefont {Schwab}},\ }\href
  {http://pubs.acs.org/doi/abs/10.1021/nl062278g} {\bibfield  {journal}
  {\bibinfo  {journal} {Nano Lett.}\ }\textbf {\bibinfo {volume} {7}},\
  \bibinfo {pages} {120} (\bibinfo {year} {2007})}\BibitemShut {NoStop}%
\bibitem [{\citenamefont {Bargatin}\ \emph {et~al.}(2012)\citenamefont
  {Bargatin}, \citenamefont {Myers}, \citenamefont {Aldridge}, \citenamefont
  {Marcoux}, \citenamefont {Brianceau}, \citenamefont {Duraffourg},
  \citenamefont {Colinet}, \citenamefont {Hentz}, \citenamefont {Andreucci},\
  and\ \citenamefont {Roukes}}]{bargatin2012large}%
  \BibitemOpen
  \bibfield  {author} {\bibinfo {author} {\bibfnamefont {I.}~\bibnamefont
  {Bargatin}}, \bibinfo {author} {\bibfnamefont {E.}~\bibnamefont {Myers}},
  \bibinfo {author} {\bibfnamefont {J.}~\bibnamefont {Aldridge}}, \bibinfo
  {author} {\bibfnamefont {C.}~\bibnamefont {Marcoux}}, \bibinfo {author}
  {\bibfnamefont {P.}~\bibnamefont {Brianceau}}, \bibinfo {author}
  {\bibfnamefont {L.}~\bibnamefont {Duraffourg}}, \bibinfo {author}
  {\bibfnamefont {E.}~\bibnamefont {Colinet}}, \bibinfo {author} {\bibfnamefont
  {S.}~\bibnamefont {Hentz}}, \bibinfo {author} {\bibfnamefont
  {P.}~\bibnamefont {Andreucci}}, \ and\ \bibinfo {author} {\bibfnamefont
  {M.}~\bibnamefont {Roukes}},\ }\href
  {http://pubs.acs.org/doi/abs/10.1021/nl2037479} {\bibfield  {journal}
  {\bibinfo  {journal} {Nano Lett.}\ }\textbf {\bibinfo {volume} {12}},\
  \bibinfo {pages} {1269} (\bibinfo {year} {2012})}\BibitemShut {NoStop}%
\bibitem [{\citenamefont {Harris}(2008)}]{harris2008electromagnetically}%
  \BibitemOpen
  \bibfield  {author} {\bibinfo {author} {\bibfnamefont {S.~E.}\ \bibnamefont
  {Harris}},\ }\href
  {http://scitation.aip.org/content/aip/magazine/physicstoday/article/50/7/10.1063/1.881806}
  {\bibfield  {journal} {\bibinfo  {journal} {Phys. Today}\ }\textbf {\bibinfo
  {volume} {50}},\ \bibinfo {pages} {36} (\bibinfo {year} {2008})}\BibitemShut
  {NoStop}%
\bibitem [{\citenamefont {Arimondo}(2006)}]{arimondo1996coherent}%
  \BibitemOpen
  \bibfield  {author} {\bibinfo {author} {\bibfnamefont {E.}~\bibnamefont
  {Arimondo}},\ }\enquote {\bibinfo {title} {Coherent population trapping in
  laser spectroscopy},}\ in\ \href
  {https://www.df.unipi.it/gruppi/arimondo/documents/EA157.pdf} {\emph
  {\bibinfo {booktitle} {Progress in Optics XXXV}}}\ (\bibinfo  {publisher}
  {Elsevier Science},\ \bibinfo {year} {2006})\ pp.\ \bibinfo {pages}
  {257--354}\BibitemShut {NoStop}%
\bibitem [{\citenamefont {Weis}\ \emph {et~al.}(2010)\citenamefont {Weis},
  \citenamefont {Rivi{\`e}re}, \citenamefont {Del{\'e}glise}, \citenamefont
  {Gavartin}, \citenamefont {Arcizet}, \citenamefont {Schliesser},\ and\
  \citenamefont {Kippenberg}}]{weis2010optomechanically}%
  \BibitemOpen
  \bibfield  {author} {\bibinfo {author} {\bibfnamefont {S.}~\bibnamefont
  {Weis}}, \bibinfo {author} {\bibfnamefont {R.}~\bibnamefont {Rivi{\`e}re}},
  \bibinfo {author} {\bibfnamefont {S.}~\bibnamefont {Del{\'e}glise}}, \bibinfo
  {author} {\bibfnamefont {E.}~\bibnamefont {Gavartin}}, \bibinfo {author}
  {\bibfnamefont {O.}~\bibnamefont {Arcizet}}, \bibinfo {author} {\bibfnamefont
  {A.}~\bibnamefont {Schliesser}}, \ and\ \bibinfo {author} {\bibfnamefont
  {T.~J.}\ \bibnamefont {Kippenberg}},\ }\href
  {http://www.sciencemag.org/content/330/6010/1520.short} {\bibfield  {journal}
  {\bibinfo  {journal} {Science}\ }\textbf {\bibinfo {volume} {330}},\ \bibinfo
  {pages} {1520} (\bibinfo {year} {2010})}\BibitemShut {NoStop}%
\bibitem [{\citenamefont {Dong}\ \emph {et~al.}(2012)\citenamefont {Dong},
  \citenamefont {Fiore}, \citenamefont {Kuzyk},\ and\ \citenamefont
  {Wang}}]{dong2012optomechanical}%
  \BibitemOpen
  \bibfield  {author} {\bibinfo {author} {\bibfnamefont {C.}~\bibnamefont
  {Dong}}, \bibinfo {author} {\bibfnamefont {V.}~\bibnamefont {Fiore}},
  \bibinfo {author} {\bibfnamefont {M.~C.}\ \bibnamefont {Kuzyk}}, \ and\
  \bibinfo {author} {\bibfnamefont {H.}~\bibnamefont {Wang}},\ }\href
  {http://www.sciencemag.org/content/338/6114/1609.short} {\bibfield  {journal}
  {\bibinfo  {journal} {Science}\ }\textbf {\bibinfo {volume} {338}},\ \bibinfo
  {pages} {1609} (\bibinfo {year} {2012})}\BibitemShut {NoStop}%
\bibitem [{\citenamefont {Wang}\ and\ \citenamefont
  {Clerk}(2012{\natexlab{a}})}]{wang2012usinginterference}%
  \BibitemOpen
  \bibfield  {author} {\bibinfo {author} {\bibfnamefont {Y.-D.}\ \bibnamefont
  {Wang}}\ and\ \bibinfo {author} {\bibfnamefont {A.~A.}\ \bibnamefont
  {Clerk}},\ }\href
  {http://journals.aps.org/prl/abstract/10.1103/PhysRevLett.108.153603}
  {\bibfield  {journal} {\bibinfo  {journal} {Phys. Rev. Lett.}\ }\textbf
  {\bibinfo {volume} {108}},\ \bibinfo {pages} {153603} (\bibinfo {year}
  {2012}{\natexlab{a}})}\BibitemShut {NoStop}%
\bibitem [{\citenamefont {Wilson}\ \emph {et~al.}(2015)\citenamefont {Wilson},
  \citenamefont {Sudhir}, \citenamefont {Piro}, \citenamefont {Schilling},
  \citenamefont {Ghadimi},\ and\ \citenamefont
  {Kippenberg}}]{wilson2015measurement}%
  \BibitemOpen
  \bibfield  {author} {\bibinfo {author} {\bibfnamefont {D.}~\bibnamefont
  {Wilson}}, \bibinfo {author} {\bibfnamefont {V.}~\bibnamefont {Sudhir}},
  \bibinfo {author} {\bibfnamefont {N.}~\bibnamefont {Piro}}, \bibinfo {author}
  {\bibfnamefont {R.}~\bibnamefont {Schilling}}, \bibinfo {author}
  {\bibfnamefont {A.}~\bibnamefont {Ghadimi}}, \ and\ \bibinfo {author}
  {\bibfnamefont {T.~J.}\ \bibnamefont {Kippenberg}},\ }\href
  {http://www.nature.com/nature/journal/v524/n7565/abs/nature14672.html}
  {\bibfield  {journal} {\bibinfo  {journal} {Nature}\ }\textbf {\bibinfo
  {volume} {524}},\ \bibinfo {pages} {325} (\bibinfo {year}
  {2015})}\BibitemShut {NoStop}%
\bibitem [{\citenamefont {Azak}\ \emph {et~al.}(2007)\citenamefont {Azak},
  \citenamefont {Shagam}, \citenamefont {Karabacak}, \citenamefont {Ekinci},
  \citenamefont {Kim},\ and\ \citenamefont {Jang}}]{azak2007nanomechanical}%
  \BibitemOpen
  \bibfield  {author} {\bibinfo {author} {\bibfnamefont {N.}~\bibnamefont
  {Azak}}, \bibinfo {author} {\bibfnamefont {M.}~\bibnamefont {Shagam}},
  \bibinfo {author} {\bibfnamefont {D.}~\bibnamefont {Karabacak}}, \bibinfo
  {author} {\bibfnamefont {K.}~\bibnamefont {Ekinci}}, \bibinfo {author}
  {\bibfnamefont {D.}~\bibnamefont {Kim}}, \ and\ \bibinfo {author}
  {\bibfnamefont {D.}~\bibnamefont {Jang}},\ }\href
  {http://scitation.aip.org/content/aip/journal/apl/91/9/10.1063/1.2776981}
  {\bibfield  {journal} {\bibinfo  {journal} {App. Phys. Lett.}\ }\textbf
  {\bibinfo {volume} {91}},\ \bibinfo {pages} {093112} (\bibinfo {year}
  {2007})}\BibitemShut {NoStop}%
\bibitem [{\citenamefont {Tian}(2012)}]{tian2012adiabatic}%
  \BibitemOpen
  \bibfield  {author} {\bibinfo {author} {\bibfnamefont {L.}~\bibnamefont
  {Tian}},\ }\href
  {http://journals.aps.org/prl/abstract/10.1103/PhysRevLett.108.153604}
  {\bibfield  {journal} {\bibinfo  {journal} {Phys. Rev. Lett.}\ }\textbf
  {\bibinfo {volume} {108}},\ \bibinfo {pages} {153604} (\bibinfo {year}
  {2012})}\BibitemShut {NoStop}%
\bibitem [{\citenamefont {Wang}\ and\ \citenamefont
  {Clerk}(2012{\natexlab{b}})}]{wang2012usingdarkmodes}%
  \BibitemOpen
  \bibfield  {author} {\bibinfo {author} {\bibfnamefont {Y.-D.}\ \bibnamefont
  {Wang}}\ and\ \bibinfo {author} {\bibfnamefont {A.~A.}\ \bibnamefont
  {Clerk}},\ }\href
  {http://iopscience.iop.org/article/10.1088/1367-2630/14/10/105010/meta}
  {\bibfield  {journal} {\bibinfo  {journal} {New J. Phys.}\ }\textbf {\bibinfo
  {volume} {14}},\ \bibinfo {pages} {105010} (\bibinfo {year}
  {2012}{\natexlab{b}})}\BibitemShut {NoStop}%
\bibitem [{\citenamefont {Gavartin}\ \emph {et~al.}(2012)\citenamefont
  {Gavartin}, \citenamefont {Verlot},\ and\ \citenamefont
  {Kippenberg}}]{gavartin2012hybrid}%
  \BibitemOpen
  \bibfield  {author} {\bibinfo {author} {\bibfnamefont {E.}~\bibnamefont
  {Gavartin}}, \bibinfo {author} {\bibfnamefont {P.}~\bibnamefont {Verlot}}, \
  and\ \bibinfo {author} {\bibfnamefont {T.~J.}\ \bibnamefont {Kippenberg}},\
  }\href {http://www.nature.com/articles/nnano.2012.97} {\bibfield  {journal}
  {\bibinfo  {journal} {Nat. Nano.}\ }\textbf {\bibinfo {volume} {7}},\
  \bibinfo {pages} {509} (\bibinfo {year} {2012})}\BibitemShut {NoStop}%
\bibitem [{\citenamefont {Bagci}\ \emph {et~al.}(2014)\citenamefont {Bagci},
  \citenamefont {Simonsen}, \citenamefont {Schmid}, \citenamefont {Villanueva},
  \citenamefont {Zeuthen}, \citenamefont {Appel}, \citenamefont {Taylor},
  \citenamefont {S{\o}rensen}, \citenamefont {Usami}, \citenamefont
  {Schliesser} \emph {et~al.}}]{bagci2014optical}%
  \BibitemOpen
  \bibfield  {author} {\bibinfo {author} {\bibfnamefont {T.}~\bibnamefont
  {Bagci}}, \bibinfo {author} {\bibfnamefont {A.}~\bibnamefont {Simonsen}},
  \bibinfo {author} {\bibfnamefont {S.}~\bibnamefont {Schmid}}, \bibinfo
  {author} {\bibfnamefont {L.~G.}\ \bibnamefont {Villanueva}}, \bibinfo
  {author} {\bibfnamefont {E.}~\bibnamefont {Zeuthen}}, \bibinfo {author}
  {\bibfnamefont {J.}~\bibnamefont {Appel}}, \bibinfo {author} {\bibfnamefont
  {J.~M.}\ \bibnamefont {Taylor}}, \bibinfo {author} {\bibfnamefont
  {A.}~\bibnamefont {S{\o}rensen}}, \bibinfo {author} {\bibfnamefont
  {K.}~\bibnamefont {Usami}}, \bibinfo {author} {\bibfnamefont
  {A.}~\bibnamefont {Schliesser}},  \emph {et~al.},\ }\href
  {http://www.nature.com/nature/journal/v507/n7490/abs/nature13029.html}
  {\bibfield  {journal} {\bibinfo  {journal} {Nature}\ }\textbf {\bibinfo
  {volume} {507}},\ \bibinfo {pages} {81} (\bibinfo {year} {2014})}\BibitemShut
  {NoStop}%
\bibitem [{\citenamefont {Regal}\ and\ \citenamefont
  {Lehnert}(2011)}]{regal2011cavity}%
  \BibitemOpen
  \bibfield  {author} {\bibinfo {author} {\bibfnamefont {C.}~\bibnamefont
  {Regal}}\ and\ \bibinfo {author} {\bibfnamefont {K.}~\bibnamefont
  {Lehnert}},\ }in\ \href
  {http://iopscience.iop.org/article/10.1088/1742-6596/264/1/012025/meta}
  {\emph {\bibinfo {booktitle} {J. of Phys.: Conf. Series}}},\ Vol.\ \bibinfo
  {volume} {264}\ (\bibinfo {organization} {IOP Publishing},\ \bibinfo {year}
  {2011})\ p.\ \bibinfo {pages} {012025}\BibitemShut {NoStop}%
\bibitem [{\citenamefont {Andrews}\ \emph {et~al.}(2014)\citenamefont
  {Andrews}, \citenamefont {Peterson}, \citenamefont {Purdy}, \citenamefont
  {Cicak}, \citenamefont {Simmonds}, \citenamefont {Regal},\ and\ \citenamefont
  {Lehnert}}]{andrews2014bidirectional}%
  \BibitemOpen
  \bibfield  {author} {\bibinfo {author} {\bibfnamefont {R.}~\bibnamefont
  {Andrews}}, \bibinfo {author} {\bibfnamefont {R.}~\bibnamefont {Peterson}},
  \bibinfo {author} {\bibfnamefont {T.}~\bibnamefont {Purdy}}, \bibinfo
  {author} {\bibfnamefont {K.}~\bibnamefont {Cicak}}, \bibinfo {author}
  {\bibfnamefont {R.}~\bibnamefont {Simmonds}}, \bibinfo {author}
  {\bibfnamefont {C.}~\bibnamefont {Regal}}, \ and\ \bibinfo {author}
  {\bibfnamefont {K.}~\bibnamefont {Lehnert}},\ }\href
  {http://www.nature.com/nphys/journal/v10/n4/abs/nphys2911.html} {\bibfield
  {journal} {\bibinfo  {journal} {Nat. Phys.}\ }\textbf {\bibinfo {volume}
  {10}},\ \bibinfo {pages} {321} (\bibinfo {year} {2014})}\BibitemShut
  {NoStop}%
\bibitem [{\citenamefont {Bochmann}\ \emph {et~al.}(2013)\citenamefont
  {Bochmann}, \citenamefont {Vainsencher}, \citenamefont {Awschalom},\ and\
  \citenamefont {Cleland}}]{bochmann2013nanomechanical}%
  \BibitemOpen
  \bibfield  {author} {\bibinfo {author} {\bibfnamefont {J.}~\bibnamefont
  {Bochmann}}, \bibinfo {author} {\bibfnamefont {A.}~\bibnamefont
  {Vainsencher}}, \bibinfo {author} {\bibfnamefont {D.~D.}\ \bibnamefont
  {Awschalom}}, \ and\ \bibinfo {author} {\bibfnamefont {A.~N.}\ \bibnamefont
  {Cleland}},\ }\href
  {http://www.nature.com/nphys/journal/v9/n11/abs/nphys2748.html} {\bibfield
  {journal} {\bibinfo  {journal} {Nat. Phys.}\ }\textbf {\bibinfo {volume}
  {9}},\ \bibinfo {pages} {712} (\bibinfo {year} {2013})}\BibitemShut {NoStop}%
\end{thebibliography}
